\documentclass[reprint, superscriptaddress, secnumarabic, amssymb, nobibnotes, aps, prl]{revtex4-1}

\usepackage{chemformula,siunitx}
\usepackage{lastpage}
\usepackage{booktabs}
\usepackage[protrusion=true, expansion=true]{microtype}
\usepackage{wrapfig}
\usepackage{multirow}
\usepackage{array}
\usepackage{ragged2e}
\justifying

\usepackage{epstopdf}
\usepackage[T1]{fontenc}
\usepackage{amsbsy}
\usepackage[T1]{fontenc}
\usepackage{amsmath}
\usepackage{amssymb}
\usepackage{bbm}
\usepackage{braket}
\usepackage{xcolor}
\allowdisplaybreaks
\usepackage{graphicx,makecell}
\usepackage[para,flushleft]{threeparttable}

\usepackage[normalem]{ulem}

\renewcommand{\approx}{\simeq}

\usepackage[colorlinks=true]{hyperref}
\hypersetup{
    unicode=false,          
    pdftoolbar=true,        
    pdfmenubar=true,        
    pdffitwindow=false,     
    pdfstartview={FitH},    
    pdftitle={Pb{2-x}Bi{x}Pd},    
    pdfauthor={Shashank Srivastava},      
    pdfsubject={},   
    pdfcreator={},   
    pdfproducer={}, 
    pdfkeywords={} {} {}, 
    pdfnewwindow=true,      
    colorlinks=true,       
    linkcolor=blue, 
    citecolor=blue,        
    filecolor=magenta,      
    urlcolor=blue          
}

\begin{document}

\title{Superconductivity in \texorpdfstring{\ch{Pb_{2-x}Bi_{x}Pd}}{Pb{2-x}Bi{x}Pd} Single Crystals}

\author{S. Srivastava}
\affiliation{Department of Physics, Indian Institute of Science Education and Research Bhopal, Bhopal, 462066, India}
\author{R.~P.~Singh}
\email[]{rpsingh@iiserb.ac.in}
\affiliation{Department of Physics, Indian Institute of Science Education and Research Bhopal, Bhopal, 462066, India}

\begin{abstract}
Chemical substitution provides an effective route to tune the structural and superconducting properties and to explore the evolution of superconductivity across related materials. In this work, we investigate the effect of Pb/Bi mixing on superconductivity in possible topological superconductors \ch{Pb2Pd} and $\beta$-\ch{Bi2Pd}, by synthesizing and characterizing single crystals of \ch{Pb_{2-x}Bi_{x}Pd} ($0.2\le x\le1.8$). Increasing Bi content triggers a structural phase transition from non-symmorphic ($I4/mcm$, $x<1$) to layered ($I4/mmm$, $x\ge1$) and induces a monotonic expansion of the lattice parameters. Across the doped range, all superconducting samples behave as weakly-coupled type-II superconductors exhibiting isotropic s-wave superconducting gaps.
\end{abstract}
\maketitle

\section{Introduction}
Superconducting ground states are highly sensitive to external perturbations such as pressure, chemical substitution, and lattice distortions, making these systems an excellent platform for investigating the relationship between crystal structure and superconductivity \cite{gui2021,attfield2011}. Among these approaches, chemical substitution has emerged as an effective and versatile tool for tuning the structural, electronic, and superconducting properties of materials  \cite{lin2022}. By modifying the crystal lattice, electronic environment, and scattering mechanisms, substitution can significantly influence the superconducting transition temperature, critical fields, and other fundamental superconducting parameters, while also providing valuable insight into the evolution of the superconducting state. Consequently, systematic compositional studies have become an important strategy for understanding the mechanisms governing superconductivity and for discovering materials with enhanced or novel superconducting properties.\\

Strong spin–orbit coupling (SOC) and structural transitions can significantly influence the superconducting state, particularly in materials with non-trivial electronic band structures \cite{Sato,nayak2008topo,qi2011tis}. It is well-established that superconducting compounds comprising high atomic number elements ($Z$) have high-SOC because the strength of SOC is directly proportional to $Z^4$. Superconducting systems with strong SOC offer an opportunity to realize unconventional properties such as violation of the Pauli limiting field \cite{kimura2007pauli,4hnbse2,uchida2019pauli}, multigap and multiband superconducting state \cite{bi2pd2015prb}, non-trivial band topology \cite{bi2pd2017scibull,Sajilesh2024hfrhge,pb2pd2022awana}.\\

\begin{figure*}[t!]
\centering
\includegraphics[width=0.9\textwidth]{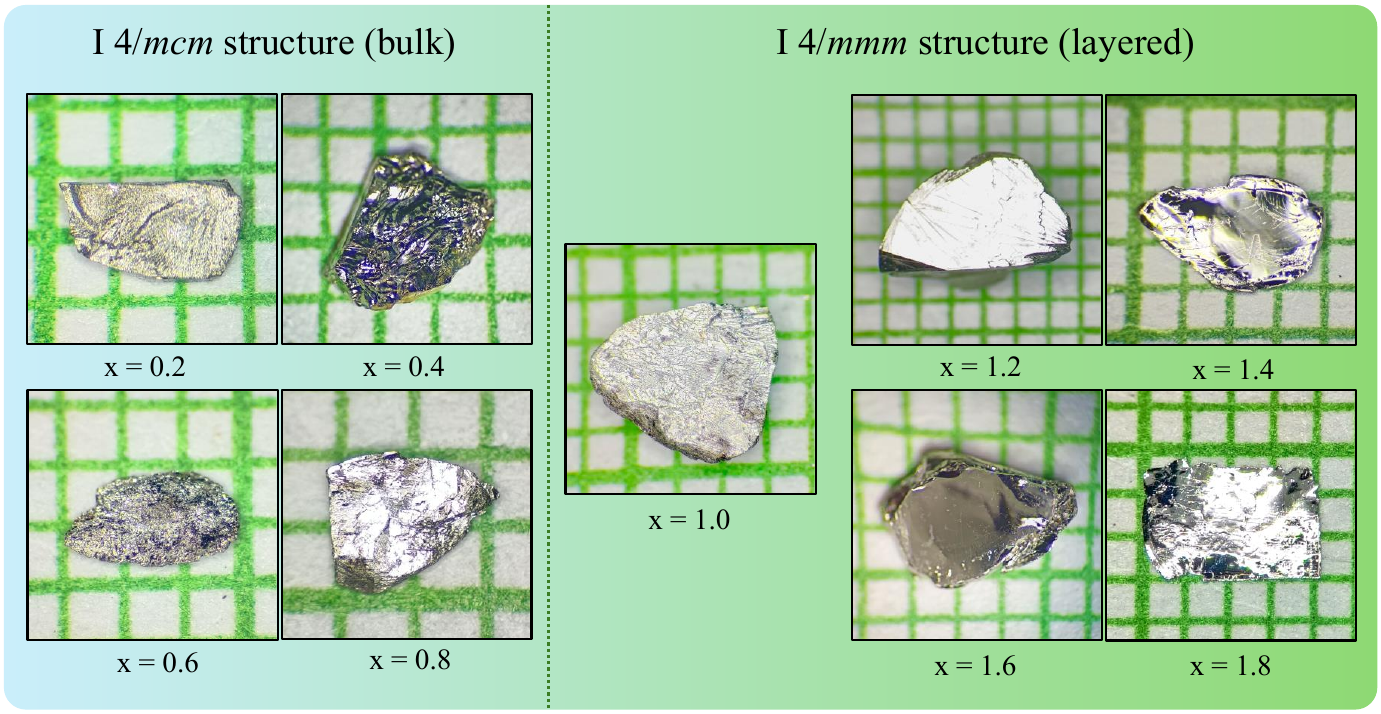}
\caption {\label{crystal} Microscopic images of as grown single crystals for \ch{Pb_{2-x}Bi_{x}Pd}. Crystals with $I4/mcm$ structure ($x=0.2-0.8$) have bulk morphology, whereas the $I4/mmm$ structure crystals ($x=1.0-1.8$) have layered structure.}
\end{figure*}

Lately, superconductivity in compounds based on high-SOC elements Pb and Bi has gained significant interest because of their non-trivial topological features \cite{tu2019ypb3,zhang2019rhpb2,ptpb3bi,xu2021superconductivity,pb2pd2022awana,sun2015alphaBiPd,benia2016alphaBiPd,sharma2024gammabipd,dimitri2018alphaBi2Pd,sakano2015Bi2Pd,NiBi3,CaBi3SrBi3}. Due to their similar atomic size and chemical properties, the doping effect of Pb/Bi on the Bi/Pb site is fascinating. Bi doping at the Pb site can induce superconductivity following dome-shaped behavior \cite{BaPbO3}, whereas various compounds have seen an increase in the superconducting critical temperature ($T_C$) in Pb doping at the Bi site \cite{csbi4te6,La4Bi3,BSCCO}. The limited number of studies on the effect of doping high-SOC elements on superconductivity underscores the need for further investigation. Moreover, doping high-SOC elements in topological materials may cause exotic modifications to the band topology, potentially providing a better understanding of topological superconductivity. Therefore, we have chosen two topologically non-trivial materials \ch{Pb2Pd} and $\beta$-\ch{Bi2Pd} to investigate the effect of mixing the high-SOC elements Pb and Bi on the superconducting, topological, and structural properties.

Both \ch{Pb2Pd} and $\beta$-\ch{Bi2Pd} crystallize in a body-centered tetragonal structure, where \ch{Pb2Pd} hosts a nonsymmorphic symmetry and $\beta$-\ch{Bi2Pd} has a layered morphology \cite{Pb2Pd,bi2pd2017scibull}. Theoretical studies on \ch{Pb2Pd} have predicted a non-trivial band topology, whereas multiple reports suggest that $\beta$-\ch{Bi2Pd} is a candidate for topological superconductivity \cite{pb2pd2022awana,bi2pd2017scibull,bi2pd2019scibull,bi2pd2020prb,bi2pd2022monolayer,bi2pd2022nature}. \ch{Pb2Pd} exhibits conventional isotropic s-wave superconductivity with a superconducting transition temperature ($T_C$) of approximately 3 \si{K}. However, Chamorro \textit{et al.} proposed that \ch{Pb2Pd} is a ${\mathbb Z}_2$ topological metal with possible topological superconductivity \cite{pb2pd2022tsc}. In contrast, $\beta$-\ch{Bi2Pd} is a superconductor with a higher $T_C$ of about 5.3 \si{K} and has attracted considerable interest because the nature of its superconducting gap and pairing mechanism remains under debate. The synthesis of single crystals of these compounds will provide insight into the change in structural morphology, which remains obscure in studies on polycrystals, and a detailed analysis of superconducting properties across the entire doping range is still lacking \cite{pb2pd1962,Pb2Pd,pb2pd2022awana,bi2pd2019science,Bi2Pd2023prl,bi2pd2015prb,Bi2Pdsingle,Bi2Pd,bi2pd2019scibull,pbdopedbi2pd,bidopedpb2pd,Biswas2016PdBi2}.

In this paper, we explore the normal and superconducting states of the single crystals of \ch{Pb_{2-x}Bi_{x}Pd}, $0.2\le x\le1.8$, with a detailed analysis for $x=0.2,0.4,0.6,$ and $1.8$ doping values. Our in-depth study includes the AC electrical resistivity, DC magnetization, and specific heat measurements to obtain the superconducting phase diagram for the single crystals. Notably, we observed a controlled structural phase transition in doping, with all samples exhibiting type-II weakly coupled superconductivity, despite \ch{Pb2Pd} being a type-I superconductor.

\section{Experimental details}
The single crystals of \ch{Pb_{2-x}Bi_{x}Pd}, $0.2\le x\le1.8$, were synthesized using the modified Bridgman technique. The high-purity elemental pieces of Pd (4N), Pb (5N), and Bi (4N) were taken in the stoichiometric ratio, respectively. The raw elements were vacuum-sealed inside conical-end quartz ampules and kept in the box furnace, as described in Figure S1 of the Supplementary Information. The ampules were heated to $873-1273$ \si{K} for various levels of doping, dwelt for $2-3$ days, and slowly cooled, followed by ice quenching. The detailed growth procedure for all single crystals is mentioned in Table S1 (see Supplementary Information). The single crystals obtained are shiny, rough, and brittle for $0.2\le x\le0.8$, like \ch{Pb2Pd}, and for $1.0\le x\le 1.8$, layered, and easily exfoliable with a metallic luster similar to \ch{Bi2Pd}, as shown in Figure \ref{crystal}.
A PANalytical X'Pert diffractometer with Cu-K$_\alpha$ radiation acquired the room-temperature powder X-ray diffraction (XRD) data for both ground and as-grown single crystals. A Photonic–Science Laue camera system was utilized to record the Laue diffraction pattern to determine the orientation and quality of a single crystal. A scanning electron microscope (SEM) was used for energy-dispersive X-ray (EDX) analysis. The vibrating sample magnetometer (VSM) mode of the 7T Quantum Design MPMS-3 superconducting quantum interference device (SQUID) magnetometer was used to measure magnetization. The 9T physical property measurement system (PPMS) determined the specific heat and resistivity using the two-tau model and the four-probe approach.

\begin{figure*}[ht!]
\includegraphics[width=2.1\columnwidth]{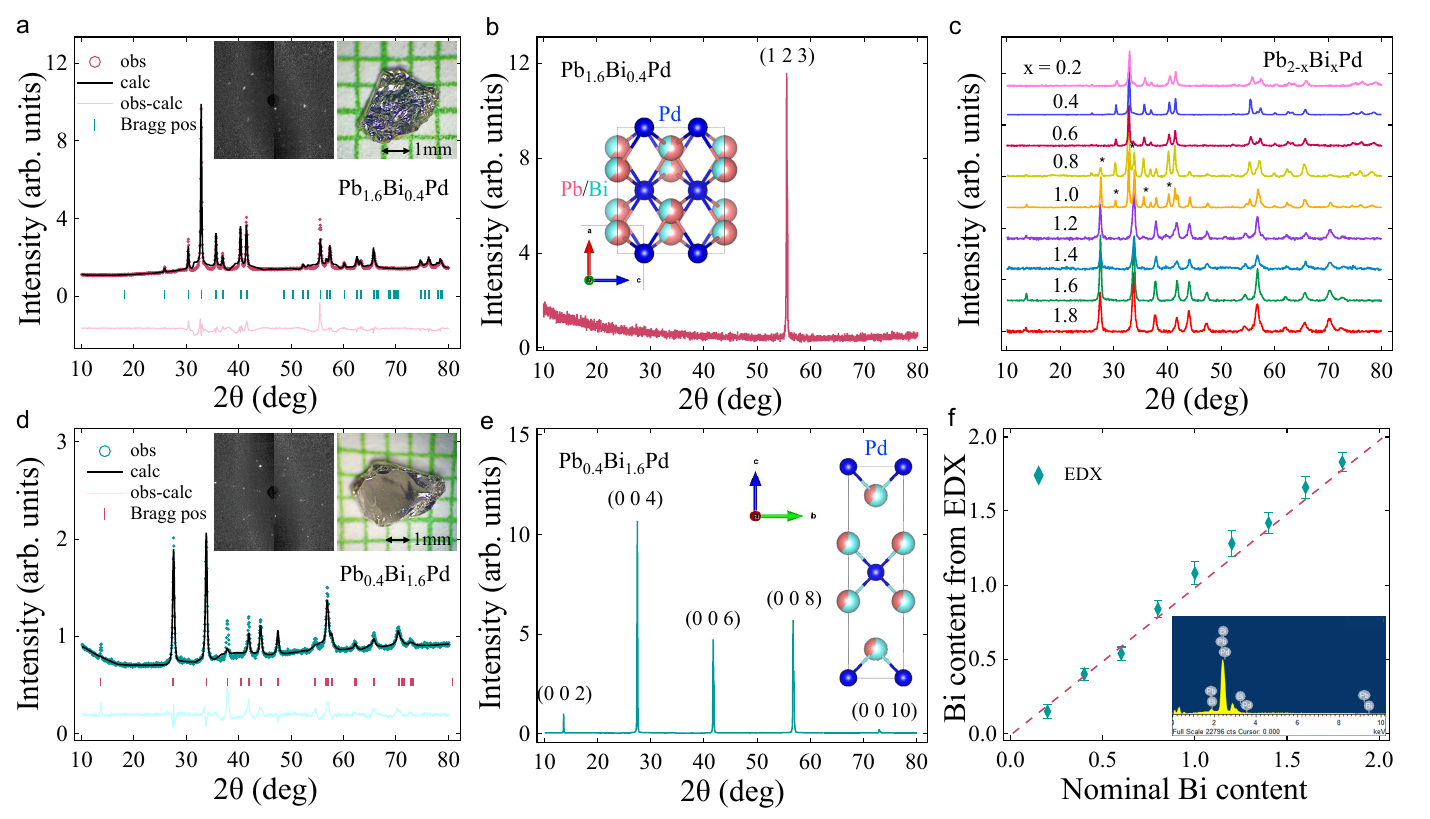}
\caption {\label{Fig1}(a) and (d) Rietveld refinement of powder XRD for \ch{Pb_{1.6}Bi_{0.4}Pd} and \ch{Pb_{0.4}Bi_{1.6}Pd} samples, respectively, at room temperature. The linked insets show the Laue diffraction patterns (left) and microscopic images (right) for the respective single crystals. (b) and (e) Room temperature XRD on the single crystals of \ch{Pb_{1.6}Bi_{0.4}Pd} and \ch{Pb_{0.4}Bi_{1.6}Pd}, respectively. The insets of (b) and (e) represent the crystal structures of \ch{Pb_{2-x}Bi_{x}Pd}, for $x<1$ and $x\ge1$, respectively. The blue balls represent the Pd atoms, and the pink/cyan ball denotes the Pb/Bi atoms. (c) Powder XRD patterns for \ch{Pb_{2-x}Bi_{x}Pd}, $0.2\le x\le1.8$. The extra peaks due to phase mixing are represented by the asterisk (*). (f) Average Bi content from EDX measurement versus the nominal Bi content in the samples.}
\end{figure*}

\section{Results and discussion}
\subsection{Sample characterization}
Rietveld refinement of the XRD patterns for the ground powder of single crystals was performed to assess phase purity using FullProf Suite software \cite{fullprof}. The single crystals were found to crystallize in the centrosymmetric body-centered tetragonal structure, with two distinct space groups (see the insets of Figure \ref{Fig1}b,e). The single crystals of \ch{Pb_{2-x}Bi_{x}Pd} with $x<1$ fit well with the non-symmorphic space group $I4/mcm$ (No. 139), while the samples with $x\ge1$ stabilize in the space group $I4/mmm$ (No. 140). Figure \ref{Fig1}c displays the XRD patterns for all compounds. The asterisk shows the extra peaks of the space group no. 140 and 139, for the single crystals with $x=0.8$ and $1$, respectively. The lattice parameters and cell volume for single crystals are summarized in Table \ref{tbl: parameters}. Figures \ref{Fig1}a,d show the refined XRD patterns for the $x=0.4$ and $1.6$ samples, respectively. The insets in the left depict the Laue diffraction patterns for the corresponding single crystals to ensure the high quality of the single crystals. All measurements were performed with the magnetic field applied along the out-of-plane direction ($H\parallel c$) for the layered crystals to compare them with \ch{Bi2Pd} under the same conditions. The quality of the single crystal was further ensured by XRD in the single crystals (Figure \ref{Fig1}b,e). EDX measurements in each sample confirm homogeneity and the required doping concentration, as shown in Figure \ref{Fig1}f, where the average Bi content is plotted against the nominal composition.

\subsection{Electrical resistivity}
\begin{figure*}[t]
\includegraphics[width=2.1\columnwidth]{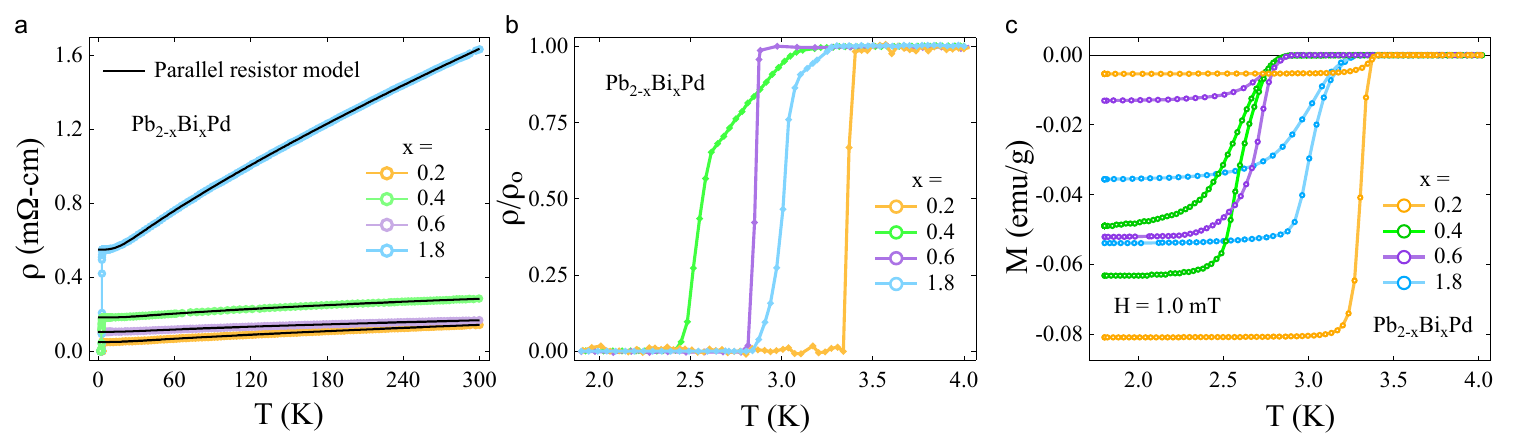}
\caption {\label{fig:2}(a) The variation of resistivity with temperature (1.9 \si{K} to 300 \si{K}) under zero magnetic field for \ch{Pb_{2-x}Bi_{x}Pd}, $x=0.2,0.4,0.6,$ and $1.8$, fitted with the parallel resistor model. (b) The normalized resistivity versus temperature indicates a zero drop in resistivity for the compounds. (c) Temperature dependence of the magnetization in ZFCW and FCC modes under 1 \si{mT} magnetic field for the compounds.}
\end{figure*}

AC electrical resistivity versus temperature between 1.9 and 300 \si{K} for samples with $x=0.2,0.4,0.6,$ and $1.8$ under zero applied field is shown in Figure \ref{fig:2}a. The normal-state resistivity decreased with temperature, indicating the metallic nature of the compounds. The ratio $\rho(300)$/$\rho(10)$ denotes the residual resistivity ratio (RRR), which lies between 1.5--3.0 (Table \ref{tbl: parameters}) for all single crystals, indicating the increase in disorder with doping in the parent compounds. The zero drop in resistivity confirms the superconductivity in $x=0.2,0.4,0.6,0.8,1.0$ and $1.8$ single crystals. Compounds with $x=1.2,1.4$ and $1.6$ do not show a zero drop due to low superconducting volume fraction (SVF) discussed in the next subsection. Figure \ref{fig:2}b and S2a (Supplementary Information) present an expanded plot of the variation in normalized resistivity with temperature, clearly showing the transition for all the mentioned compounds. The onset of zero drop in resistivity at 3.37, 3.15, 2.88, 3.67, 3.24 and 3.30 \si{K} indicates superconducting transition temperature ($T_C$) for $x=0.2,0.4,0.6,0.8,1.0$ and $1.8$, respectively. Furthermore, the normal-state data for resistivity versus temperature exhibit saturating behavior, which Wiesmann described using the parallel resistor model \cite{parallel}. The data for all the measured compounds fit nicely with the parallel resistor model, which relates the temperature-dependent resistivity $\rho(T)$ with the saturated resistivity $\rho_s$ at high temperatures, and the ideal contribution of resistivity $\rho_{i}$ (Eq. \ref{Eq:Parallel}).

\begin{equation}
\frac{1}{\rho(T)}= \frac{1}{\rho_{s}}+ \frac{1}{\rho_{i}(T)},
\label{Eq:Parallel}
\end{equation}
where, $\rho_{i}(T)$ = $\rho_{i,0}$ + $\rho_{i,L}(T)$. $\rho_{i,0}$ denotes the temperature-invariant residual resistivity attributed to impurity scattering. $\rho_{i,L}(T)$ represents the temperature-variable resistivity caused by electron-phonon scattering, defined by Wilson as \cite{parallel2},  
\begin{equation}
{\rho_{i,L}(T)}= A_0 \left(\frac{T}{\theta_{R}}\right)^{n_0} \int_{0}^{{\theta_{R}}/T} \frac{x^{n_0}}{(e^{-x}-1)(1-e^{x})} dx.
\label{Eqn:Parallel}
\end{equation}
Depending on the type of interaction, $n_0$ typically takes the values 2, 3, or 5; $A_0$ is a material-dependent variable; and $\theta_{R}$ represents the Debye temperature \cite{parallel3}. The data for the samples mentioned were best fitted with $n_0=3$. The values of $\theta_{R}$ obtained from the fitting are listed in Table \ref{tbl: parameters}, which are close to the values reported for the parent compounds.

\subsection{Magnetization}
The variation of the magnetic moment with temperature was measured for \ch{Pb_{2-x}Bi_{x}Pd}, $0.2\le x\le1.8$, single crystals under 1 \si{mT} applied magnetic field in the zero field-cooled warming (ZFCW) and field-cooled cooling (FCC) modes. All samples were found to be superconducting (variation of $T_C$ in Figure \ref{Fig:4}), among which $x=0.2,0.4,0.6,0.8,1.0$ and $1.8$ exhibit bulk superconductivity. The normalized magnetization versus temperature of the compounds $x=1.2,1.4,1.6$ with SVF less than 2\% is also shown in Supplementary Information (Figure S2b and Figure \ref{fig:2}c). The SVF of the compounds $x=0.2,0.4,0.6,0.8,1.0,1.2,1.4,1.6$ and $1.8$ are $128,99,79,133,63,1.3,0.2,0.1$ and $110$\%, respectively. Since samples with $x=0.8$ and $1$ exhibit additional peaks, a detailed study has been performed on single crystals with $x=0.2,0.4,0.6,$ and $1.8$. Their superconducting critical temperatures, which is defined by the bifurcation of the ZFCW and FCC curves, are mentioned in Table \ref{tbl: parameters}. The division of the FCC from the ZFCW curves implies substantial flux pinning, which is the initial indication of type II superconductivity.

\begin{figure*}[ht!]
\includegraphics[width=2.1\columnwidth]{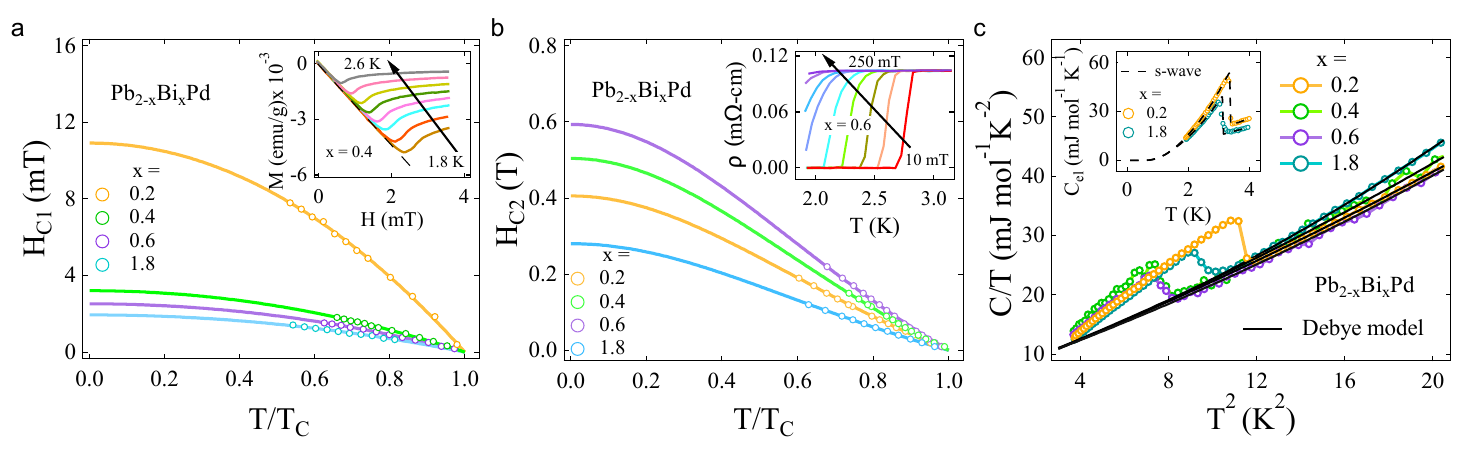}
\caption {\label{Fig:3} The reduced temperature ($T/T_C$) variation of the (a) lower critical field, and the (b) upper critical field for \ch{Pb_{2-x}Bi_{x}Pd} single crystals, fitted with the GL equations to obtain the $H_{C1}$(0) and $H_{C2}$(0), respectively. The inset of (a) shows the magnetic field variation of magnetization for \ch{Pb_{1.6}Bi_{0.4}Pd} at different temperatures (1.8 \si{K} to 2.6 \si{K}), and the inset of (b) shows the temperature variation of resistivity between different applied magnetic fields of 10 \si{mT} to 250 \si{mT} for \ch{Pb_{1.4}Bi_{0.6}Pd}. (c) $C/T$ versus $T^2$ at $H=0$~\si{mT} fitted with Debye model (Eq. \ref{Eqn:Debye}). Inset: Temperature variation of electronic specific heat, $C_{el} (T)$ best fitted with the isotropic s-wave model.}
\end{figure*}

\begin{figure}[t]
\includegraphics[width=1.02\columnwidth]{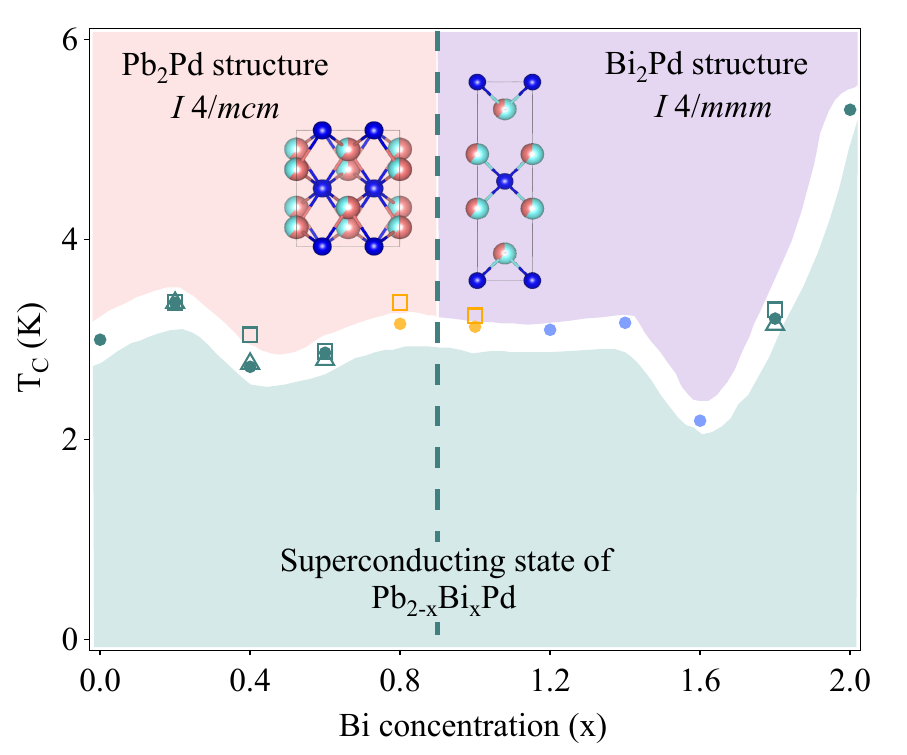}
\caption {\label{Fig:4} Superconducting phase diagram for the \ch{Pb_{2-x}Bi_{x}Pd} single crystals. The green, yellow, and blue markers indicate single-phase bulk superconductors, mixed-phase bulk superconductors, and superconductors with volume fraction less than 2\%, respectively. Circle, square, and triangle markers show $T_C$ from magnetization, resistivity, and specific heat, respectively. The green dotted line marks the transition in the crystal structure. The $T_C$ for \ch{Pb2Pd} and \ch{Bi2Pd} were taken from Refs. \cite{Pb2Pd,Bi2Pd}.}
\end{figure}

Magnetization was also measured at various temperatures below $T_C$ within a low magnetic field range for the single-phase bulk superconductors. The inset of Figure \ref{Fig:3}a depicts the low field M-H for \ch{Pb_{1.6}Bi_{0.4}Pd}. At each temperature, the lower critical field, $H_{C1}$, is determined from the divergence of the curves from the Meissner line. The Ginzburg-Landau (GL) equation (Eq. \ref{eqn:HC1}) was used to describe the variation of $H_{C1}$ with reduced temperature, as demonstrated in Figure \ref{Fig:3}a.
\begin{equation}
H_{C1}(T)=H_{C1}(0)[{1-(t)^{2}}],  \quad  \text{where} \;  t = \frac{T}{T_{C}}.
\label{eqn:HC1}
\end{equation}
The values of $H_{C1}(0)$ obtained from the fitting of all the bulk superconductors were in the range of 1-11 \si{mT}, and are listed in Table \ref{tbl: parameters}. These values are less than $H_{C1}(0)$ of \ch{Bi2Pd} \cite{Bi2Pdsingle}.

Similarly, the upper critical field values, $H_{C2}$, as a function of reduced temperature, were accurately modeled using the GL equation specified in Eq. \ref{eqn:HC2}, for all bulk superconductors (see Figure \ref{Fig:3}b). The variation in temperature of the resistivity and magnetization data can be used to obtain the values of $H_{C2}$(T). As the magnetic field increases, the resistivity (and magnetization) curves indicate a decrease in $T_C$, yielding the $H_{C2}$ values at different temperatures. The inset of Figure \ref{Fig:3}b presents the field-dependent $\rho$-T for the \ch{Pb_{1.4}Bi_{0.6}Pd} single crystal.
\begin{equation}
H_{C2}(T) = H_{C2}(0)\left[\frac{1-(t)^{2}}{1+(t)^{2}}\right],  \quad  \text{where} \;  t = \frac{T}{T_{C}}.
\label{eqn:HC2}
\end{equation}
Extrapolation of the $H_{C2}$(T) versus $T/T_C$ data gives an upper critical field $H_{C2}$(0) for the samples, and the value of $H_{C2}$(0) increased with higher doping. The values of $H_{C2}$(0) for doped compounds are less than $H_{C2}(0)$ of \ch{Bi2Pd} \cite{Bi2Pdsingle}.

There are two possible processes by which a magnetic field can destroy superconductivity. (i) Spin paramagnetic effect, the breakdown of Cooper pairs induced by the Zeeman effect, is referred to as the Pauli limiting field, $H_{C2}^{P}$(0) = 1.86 $T_{C}$ \cite{Chandrasekhar1962pauli, Clogston1962pauli}. (ii) The orbital limiting effect arises from the enhancement of the kinetic energy of the supercurrent surpassing the superconducting gap energy, which may be assessed by the Werthamer-Helfand-Hohenberg (WHH) theory for a type-II superconductor, as defined in Eq. \ref{eqn:WHH} \cite{WHH1966orbital, Helfand1966orbital}.
\begin{equation}
H^{orbital}_{C2}(0) = -\alpha T_{C} \left.{\frac{dH_{C2}(T)}{dT}}\right|_{T=T_{C}}, 
\label{eqn:WHH}
\end{equation}
Here, $\alpha$ denotes a constant called the purity factor, which assumes distinct values for dirty and clean limit superconductors (0.69 and 0.73, respectively). The values of $H_{C2}$(0), $H_{C2}^{P}$(0), and $H^{orbital}_{C2}$(0) for the dirty limit case are compared in Table \ref{tbl: parameters}. Notably, $H_{C2}$(0) is much smaller than the Pauli limiting field for each sample, indicating that orbital effects are essential in pair breaking. The influence of orbital effects can be estimated with the Maki parameter, $\alpha_{m}$ = $\sqrt{2}{H_{C2}^{orbital}(0)}/{H_{C2}^{P}(0)}$ providing $\alpha_{m}$ within the range of 0.07 to 0.1 for the limiting field values of the compounds.

The coherence length, $\xi_{GL}$, and the penetration depth, $\lambda_{GL}$, referred to as characteristic length parameters, can be estimated by substituting $H_{C1}$(0) and $H_{C2}$(0) using the following equations.
\begin{equation}
    H_{C1}(0) = \frac{\phi_{0}}{4\pi\lambda_{GL}^2}\left[ln \frac{\lambda_{GL}}{\xi_{GL}} + 0.12\right],
\label{lambda}
\end{equation}
\begin{equation}
    H_{C2}(0) = {\frac{\phi_{0}}{2\pi \xi_{GL}^2}},
\label{xi}
\end{equation}
where $\phi_0$, the magnetic flux quantum, is a constant \cite{Cava2007coherence, tinkham2004introduction}. The values of $\xi_{GL}$ and $\lambda_{GL}$ for the samples lie in the range of  235-280 \si{nm} and 1700-5100 \si{nm}, respectively, which are comparably higher than the $\xi_{GL}$ and $\lambda_{GL}$ for the \ch{Bi2Pd} \cite{Bi2Pdsingle}.

The Ginzburg-Landau (GL) parameter $\kappa_{GL}$ was defined by GL theory to distinguish between type-I and type-II superconductors. The values of $\kappa_{GL} = \frac{\lambda_{GL}}{\xi_{GL}}$ were computed for the compounds. The values increase with doping ($6<\kappa_{GL}<19$), which is much higher than $1/\sqrt{2}$, confirming that Bi-doping in \ch{Pb2Pd} changes it from a type-I to a type-II superconductor. The thermodynamic critical field parameter at 0 \si{K}, $H_C(0)$ can also be assessed using $H_{C1}$(0), $H_{C2}$(0) and $\kappa_{GL}$ with the given equation \cite{tinkham2004introduction}: 
\begin{equation}
    H_C^2(0) ln\kappa_{GL} = {H_{C1}(0) H_{C2}(0)}.
\label{thermo}
\end{equation}
The values of $H_C(0)$ were found to decrease with doping concentrations and are less than those of \ch{Bi2Pd} \cite{Bi2Pdsingle}. All the superconducting characterization parameters estimated here are outlined in Table \ref{tbl: parameters} for the \ch{Pb_{2-x}Bi_{x}Pd}, $x=0.2,0.4,0.6$, and $1.8$ single crystals compared to the parent compounds.

\begin{table*}[t]
\caption{\RaggedRight{Superconducting and normal state specifications extracted from the XRD, magnetization, resistivity, and specific heat measurements for \ch{Pb_{2-x}Bi_{x}Pd} ($0.2\le x\le1.8$) series compared with the reported values for the parent compounds \ch{Pb2Pd} and \ch{Bi2Pd}.}}
\label{tbl: parameters}
\setlength{\tabcolsep}{11.0pt}
\renewcommand{\arraystretch}{1.3} 
\begin{center}
\begin{tabular}[b]{lccccccc}\hline
Parameters                                  & unit                  & 0 \cite{Pb2Pd} & 0.2 & 0.4 & 0.6 & 1.8 & 2 \cite{Bi2Pd,Bi2Pdsingle,Biswas2016PdBi2}\\
\hline
Space group                                  &          -            & $I4/mcm$ & $I4/mcm$ & $I4/mcm$ & $I4/mcm$ & $I4/mmm$ & $I4/mmm$      \\
$a=b$                                          & \AA              & 6.8593 & 6.8687 & 6.8771 & 6.8811 & 3.3646 & 3.37       \\
$c$                                          & \AA                & 5.8383 & 5.8479 & 5.8723 & 5.8815 & 12.9441 & 12.96       \\
$V_{cell}$                                          & \AA$^3$                & 274.65 & 275.90 & 277.73 & 278.49 & 146.53 & 147.18       \\
$T_{C}$                               & \si{K}                & 3.00 & 3.37 & 2.73 & 2.88 & 3.21 & 5.30       \\
$H_{C1}(0)$                                 & \si{mT}               & 34 & 10.92 & 3.21 & 2.52 & 1.94 & 22.50       \\ 
$H_{C2}^{res}$(0)                           & \si{T}                & - & 0.429 & 0.492 & 0.592 & 0.446 & 0.610       \\
$H_{C2}^{P}$(0)                           & \si{T}                  & - & 6.20 & 5.02 & 5.30 & 6.11 & 9.75       \\
$H_{C2}^{orb}$(0)                           & \si{T}                & - & 0.339 & 0.36 & 0.362 & 0.302 & 0.734       \\
$H_{C}(0)$                              & \si{mT}                & - & 50.8 & 24.3 & 22.5 & 17.2 & 88.8       \\
$\alpha_{m}$                                  &      -              & - & 0.077 & 0.101 & 0.097 & 0.070 & 0.106       \\
$\xi_{GL}$                                  & \si{nm}               & 1817 & 277.12 & 258.77 & 235.90 & 271.78 & 232       \\
$\lambda_{GL}^{mag}$                        & \si{nm}               & 1011.2 & 1705.72 & 3783.99 & 4462.6 & 5075.39 & 1320      \\
$\kappa_{GL}$                               & -                     & 0.55 & 6.15 & 14.62 & 18.92 & 18.67 & 5.7       \\
$\rho_{300K}/\rho_{10K}$               & -                     & 19 & 2.9 & 1.56 & 1.6 & 2.96 & 2.8       \\
$\gamma_{n}$                                & \si{mJmol^{-1}K^{-2}} & 5.23 & 6.39 & 5.22 & 5.14 & 5 & -      \\
$\theta_{R}$                           & \si{K}                & 116 & 122.6 & 132.2 & 137.1 & 116.5 & 134        \\
$\theta_{D}^{sp}$                           & \si{K}                & 152 & 158.65 & 154.20 & 156.53 & 157.93 & -        \\
$\frac{\Delta(0)}{k_{B}T_{C}}$              &  -                    & 1.87 & 1.8 & 1.82 & 1.75 & 1.77 & 2.05       \\
$\lambda_{e-ph}$                            & -                     & 0.67 & 0.69 & 0.65 & 0.65 & 0.68 & -       \\
$n$                               & $10^{28}$ \si{m^{-3}}                    & 2.96 & 3.51 & 1.43 & 0.40 & 2.63 & -        \\
$m^{*}/m_{e}$                               & -                     & 2.59 & 2.93 & 3.20 & 4.8 & 2.38 & -        \\
$v_{F}$                                     & 10$^{5}$ m/s          & 4.29 & 4.01 & 2.73 & 1.19 & 4.54 & -        \\
$\xi_{0}/l_e$                                   & -               & 3.125 & 11.5 & 21.9 & 4.9 & 103 & -      \\
$T_F$                                       & \si{K}                & 13629 & 15484 & 7811 & 2232 & 15612 & 6497     \\
$T_C/T_F$                                       & -                & 0.00019 & 0.00022 & 0.00035 & 0.00129 & 0.00021 & 0.00081     \\\hline
\end{tabular}
\par\medskip\footnotesize
\end{center}
\end{table*}

\subsection{Specific heat}
The bulk superconductivity in the \ch{Pb_{2-x}Bi_{x}Pd} single crystals, with $x=0.2, 0.4, 0.6,$ and $1.8$, is validated by the temperature dependence of the specific heat measurements $C(T)$. The superconducting transition temperature $T_{C,mid}=3.37, 2.76, 2.80,$ and $3.15$ \si{K} for the $x=0.2,0.4,0.6,$ and $1.8$ samples, respectively, is manifested by the midpoint of the significant jump in $C(T)/T$ versus $T^2$ in the zero magnetic field (Figure \ref{Fig:3}c). The Debye-Sommerfeld relation (Eq. \ref{Eqn:Debye}) is used to fit the $C(T)/T$ data above $T_C$. The electronic contribution to the specific heat is represented by the first term $\gamma_nT$, while the phononic and anharmonic contributions are shown by the terms $\beta_3T^3$ and $\beta_5T^5$, respectively. The fitting of the low-temperature data with Eq. \ref{Eqn:Debye} for the measured samples is shown in Figure \ref{Fig:3}c, which provides the Sommerfeld coefficient $\gamma_n$, the Debye constant $\beta_3$ and $\beta_5$ for the measured samples.
\begin{equation}
C/T = \gamma_{n} + \beta_{3}(T^{2}) + \beta_{5}(T^{2})^2.
\label{Eqn:Debye}
\end{equation}
The following relation can be used to evaluate the density of state at the Fermi level $D_{C}(E_{F})$:
\begin{equation}
\gamma_{n} = \left(\frac{\pi^{2} k_{B}^{2}}{3}\right) D_{C}(E_{F}),
\label{Eqn:gamman}
\end{equation}
here, the Boltzmann constant, $k_{B}$ $\approx$ 1.38 $\times$ 10$^{-23}$ \si{JK^{-1}}. The Debye constant $\beta_{3}$ and the Debye temperature $\theta_{D}$ can be related via the following relation: $\theta_{D}$ = $\left(\frac{12\pi^{4} R N}{5\beta_{3}}\right)^{\frac{1}{3}}$. The values of $\theta_{D}$ are close to the $\theta_{R}$ values obtained from the resistivity data, and lie in the range of the value obtained for the parent compound \ch{Pb2Pd} \cite{Pb2Pd}. The samples \ch{Pb_{2-x}Bi_{x}Pd} have three atoms per formula unit $N$, and the value of the universal gas constant R is 8.314 \si{Jmol^{-1}K^{-1}}.

The above result can then be used to obtain the electron-phonon coupling constant $\lambda_{e-ph}$ using the inverted McMillan's equation \cite{mcmillan1968transition}, which is given as
\begin{equation}
\lambda_{e-ph} = \frac{\mu^{*}\mathrm{ln}(\theta_{D}/1.45T_{C})+1.04}{(1-0.62\mu^{*})\mathrm{ln}(\theta_{D}/1.45T_{C})-1.04};
\label{eqn2:Lambda}
\end{equation}
In the case of intermetallic compounds, the repulsive-screened Coulomb pseudopotential parameter $\mu^{*}$ is taken to be 0.13. The value of $\lambda_{e-ph}$ is between 0.65 and 0.69 of the measured variables, indicating electron-phonon coupling similar to that of the other weakly coupled superconductors, comparable to the parent compound \ch{Pb2Pd} \cite{Pb2Pd}.

To determine the nature of the superconducting gap symmetry in the sample, we assessed the normalized electronic specific heat as a function of the reduced temperature. The phononic and anharmonic contributions to total specific heat $C$ can be deduced to estimate the specific electronic heat $C_{el}$. The jump in normalized specific heat $\frac{\Delta C_{el}}{\gamma_{n}T_C}$ ranges from 1.36 to 1.53, comparable to the weak coupling limit established by BCS theory (1.43). Furthermore, $C_{el}/\gamma_{n}T$ versus reduced temperature $T/T_C$ is well fitted using the isotropic single-gap BCS model (Eq. \ref{Eq:swave}) \cite{padamsee1973quasiparticle}, as illustrated in the inset of Figure \ref{Fig:3}c. The relationship between the entropy S$_{el}$ and C$_{el}$ is C$_{el}$ = $t\frac{dS_{el}}{dt}$, where t = $\frac{T}{T_{C}}$ is the reduced temperature. The following equation relates the temperature-dependent BCS energy gap, $\Delta(t) = \tanh[1.82\{1.018(\frac{1} {t}-1)\}^{0.51}]$ to the entropy: 
\begin{equation}
\frac{S_{el}}{\gamma_{n} T_{C}}= -\frac{6}{\pi^{2}} \left(\frac{\Delta(0)}{k_{B} T_{C}}\right)\int_{0}^{\infty}[ f_{y}ln(f_{y})
+(1-f_{y})ln(1-f_{y})] dy.
\label{Eq:swave}
\end{equation}
Hence, with respect to the Fermi energy $y$ = $\xi/\Delta(0)$, the Fermi function $f_{y}(\xi)$ = $[1+e^{\frac{E(\xi)}{k_{B}T}}]^{-1}$ is integrated. The Fermi energy and the normal electron energy are related as $E(\xi)=\sqrt{\xi^{2}+\Delta^{2}(t)}$. The optimal fit of the normalized specific heat data yields a superconducting energy gap ranging from $1.75$ to $1.82$ for all samples, which is close to the weakly coupled BCS gap of 1.76. With ${\Delta C_{el}}/{\gamma_{n}T_C}$ and ${\Delta(0)}/{k_{B}T_{C}}$ around the BCS values, the single crystals of \ch{Pb_{2-x}Bi_{x}Pd} exhibit weakly coupled superconductivity. Interestingly, s-wave superconductivity persists up to the $x=1.8$ composition, despite reports of unconventional superconductivity in \ch{Bi2Pd}, indicating that partial Pb substitution does not induce observable deviations from s-wave superconductivity, inferred from the specific heat measurements.\\

The superconducting parameters listed in Table \ref{tbl: parameters} mostly exhibit non-monotonic evolution with Bi substitution, indicating that chemical substitution modifies the superconducting state by altering the electronic environment and disorder. The lower critical field and the Sommerfeld coefficient decrease with increasing Bi content, whereas the London penetration depth increases. In contrast, other superconducting parameters, including the upper critical field, coherence length, and GL parameter, exhibit a non-monotonic composition dependence. Despite these variations, all superconducting compositions are consistently described as conventional type-II s-wave superconductors, suggesting that the pairing symmetry remains robust against Bi substitution.

\subsection{Electronic parameters}
The electronic parameters for \ch{Pb_{2-x}Bi_{x}Pd} ($x=0.2,0.4,0.6,$ and $1.8$) single crystals were calculated using a series of equations (see Eqs. \ref{eqn:gf}, \ref{eqn:le,n}) relating $\gamma_n$ to the mean free path $l_{e}$ and the effective mass $m^{*}$, considering a spherical Fermi surface \cite{mayoh2017equations}. 
\begin{equation}
m^{*} = \left(\frac{3}{\pi}\right)^{2/3}\frac{\hbar^{2}N_{A}\gamma_{n}}{k_{B}^{2}V_{\mathrm{f.u.}}n^{1/3}}.
\label{eqn:gf}
\end{equation}

\begin{equation}
n = \frac{1}{3\pi^{2}}\left(\frac{m^{*}v_{\mathrm{F}}}{\hbar}\right)^{3},\quad \textit{l}_{e} = \left(\frac{3{\hbar}^{3}}{\rho_{0}}\right)\left(\frac{\pi}{em^{*}v_{\mathrm{F}}}\right)^2.
\label{eqn:le,n}
\end{equation}

$n$, $v_\mathrm{F}$, and $V_\mathrm{f.u.}$ denote electronic carrier density, Fermi velocity, and volume of the formula unit, respectively. Furthermore, dirty limit superconductors affect the GL penetration depth $\lambda_{GL}$(0), which can be expressed in terms of the London penetration depth $\lambda_{L}$ using Eq. \ref{eqn:f}.

\begin{equation}
\lambda_{GL}^2(0) = 
\left(\frac{m^*}{\mu_0
ne^{2}}\right)
\left(1+\frac{\xi_{0}}{\textit{l}_{e}}\right).
\label{eqn:f}
\end{equation}
BCS coherence length $\xi_{0}$ can also be expressed in terms of the GL coherence length $\xi_{GL}$ using the equation. 
\begin{equation}
\xi_{0} = \xi_{GL}(0)\frac{2\sqrt{3}}{\pi}\left(1+\frac{\xi_{0}}{\textit{l}_{e}}\right)^{1/2}.
\label{eqn:xil}
\end{equation}

The above Eqs. \ref{eqn:gf}-\ref{eqn:xil} were used to calculate the values of $n$, $m^*$, $\xi_0/l_e$ using the previously calculated values of $\gamma_n$, $\rho_{0}$ and $\xi_{GL}(0)$ for the \ch{Pb_{2-x}Bi_{x}Pd} single crystals. Additionally, Eq. \ref{eqn:tf} can be used to compute the Fermi temperature, $T_F$ \cite{hillier1997classification}.

\begin{equation}
k_{B}T_{F} = \frac{\hbar^{2}}{2m^{*}}{(3\pi^{2}n)^{2/3}}.
\label{eqn:tf}
\end{equation}
The calculated values of all these electronic parameters are mentioned in Table \ref{tbl: parameters}. The ratio $\xi_0/l_e$ lies in the regime of a dirty limit superconductor for all single crystals. Y. J. Uemura divided superconductors into conventional and unconventional superconductors \cite{Uemura} based on the ratio of T$_{C}$ and T$_{F}$. The unconventional range of the superconductor lies in the interval 0.01 $\leq$ T$_{C}$/T$_{F}$ $\leq$ 0.1, whereas conventional superconductors have T$_{C}$/T$_{F}$ $\geq$ 0.1. The value of $T_C/T_F$ decreased as the doping increased from $x=0.2$ to $x=0.6$, moving closer to the unconventional range, as shown in Figure S3 (see Supplementary Information).

\subsection{Superconducting phase diagram}
The structural transition and the variation in the superconducting critical temperature are summarized in the superconducting phase diagram in Figure \ref{Fig:4}. As the doping level increases, the crystal structure transforms from a non-symmorphic $I4/mcm$ structure to a layered $I4/mmm$ structure. Superconducting phase diagram was constructed using the $T_C$ values obtained independently from magnetization, electrical resistivity, and specific heat measurements, which exhibit excellent agreement within the experimental uncertainty. The bulk superconductors, mixed-phase bulk superconductors, and superconductors with SVF less than 2\% are indicated by different colors. The superconducting critical temperature shows a non-monotonic trend, with maximum and minimum values of 3.37 and 2.19 \si{K} for $x=0.2$ and $1.6$, respectively. The $T_C$ increases abruptly when the doping value goes above $x=1.6$. The phase diagram reveals that the superconducting properties remain robust against structural transition. Furthermore, our findings indicate that Pb/Bi substitution plays a surprisingly minimal role in modifying the underlying superconducting state. This contrasts with previous reports on superconducting materials, which suggest that compositional variations of high-spin orbit-coupling (SOC) elements should substantially alter the nature of superconductivity \cite{Sato,BaPbO3,csbi4te6,La4Bi3,BSCCO}.

\section{Conclusion}
This work discusses the detailed structural and superconducting properties of \ch{Pb_{2-x}Bi_{x}Pd}, $x=0.2,0.4,0.6,$ and $1.8$. The single crystals of \ch{Pb_{2-x}Bi_{x}Pd}, $0.2\le x\le1.8$, were grown using the modified Bridgman technique, and the high-quality crystalline nature was confirmed by Laue diffraction. XRD analysis of powdered single crystals determined the structural transition, where compounds with $x<1$ and $x\ge1$ encompassed the space group $I4/mcm$ and $I4/mmm$, respectively. Although \ch{Pb2Pd} is a type-I superconductor, Bi doping leads to the emergence of type-II superconductivity. Magnetization measurements reveal bulk superconductivity for compositions $x=0.2,0.4,0.6,$ and $1.8$, each crystallizing in a single structural phase. AC transport and heat-capacity measurements were performed on bulk superconductors to determine the normal- and superconducting-state parameters. The superconducting energy gap values indicate that the superconductors are weakly coupled and have an isotropic s-wave superconducting gap. The superconducting critical temperatures of the samples are consistent with all measurements, as shown in the superconducting phase diagram, indicating that the substitution of Pb/Bi plays a surprisingly minimal role in modifying the underlying superconducting state, in contrast to previous reports on similar systems. Future investigations combining detailed electronic structure calculations and microscopic experimental probes will be essential to elucidate the origin of the suppression of bulk superconductivity in the intermediate Bi-substituted compositions and robust s-wave superconductivity up to $x=1.8$ doping. However, to understand the effect of doping on the topology of the band structure and superconducting properties, we can further analyze by combining density functional theory calculations to map topology evolution across the series, muon spin rotation spectroscopy to probe microscopic ground state properties, thermal transport, and point-contact spectroscopy for high-resolution gap mapping, and angle-resolved photoemission spectroscopy to visualize Fermi surface changes. This comprehensive approach will clarify whether topological features persist in doped \ch{Pb_{2-x}Bi_{x}Pd} and determine their influence on superconductivity.

\section*{Conflicts of interest}
There are no conflicts to declare.

\section*{Data availability}
All data supporting the findings of this study are available in the article and in its Supplementary Information. Additional data is available from the corresponding author upon reasonable request.

\section*{Acknowledgements}
S.~S. acknowledges the funding agency, University Grants Commission (UGC), Government of India, for the Senior Research Fellowship (SRF). R.~P.~S. acknowledges the Anusandhan National Research Foundation (ANRF), erstwhile Science and Engineering Research Board (SERB), Government of India, for the Core Research Grant No. CRG/2023/000817.

\bibliographystyle{revtex}
\bibliography{Library}

\clearpage

\onecolumngrid

\begin{center}
    {\Large \textbf{Supporting Information for \\"Superconductivity in \texorpdfstring{\ch{Pb_{2-x}Bi_{x}Pd}}{Pb{2-x}Bi{x}Pd} Single Crystals"}}
\end{center}

\vspace{10pt}


\setcounter{figure}{0} 

\renewcommand{\thefigure}{S\arabic{figure}}  

\setcounter{table}{0} 

\renewcommand{\thetable}{S\arabic{table}}  


In the supporting information, we present the figures related to the setup for the modified Bridgman technique, the superconducting critical temperatures for the compounds having superconducting volume fraction less than 1\%, and the Uemura plot. The table summarizes single crystal growth cycle for all the crystals.

\begin{figure}[htp]
\includegraphics[width=0.62\columnwidth]{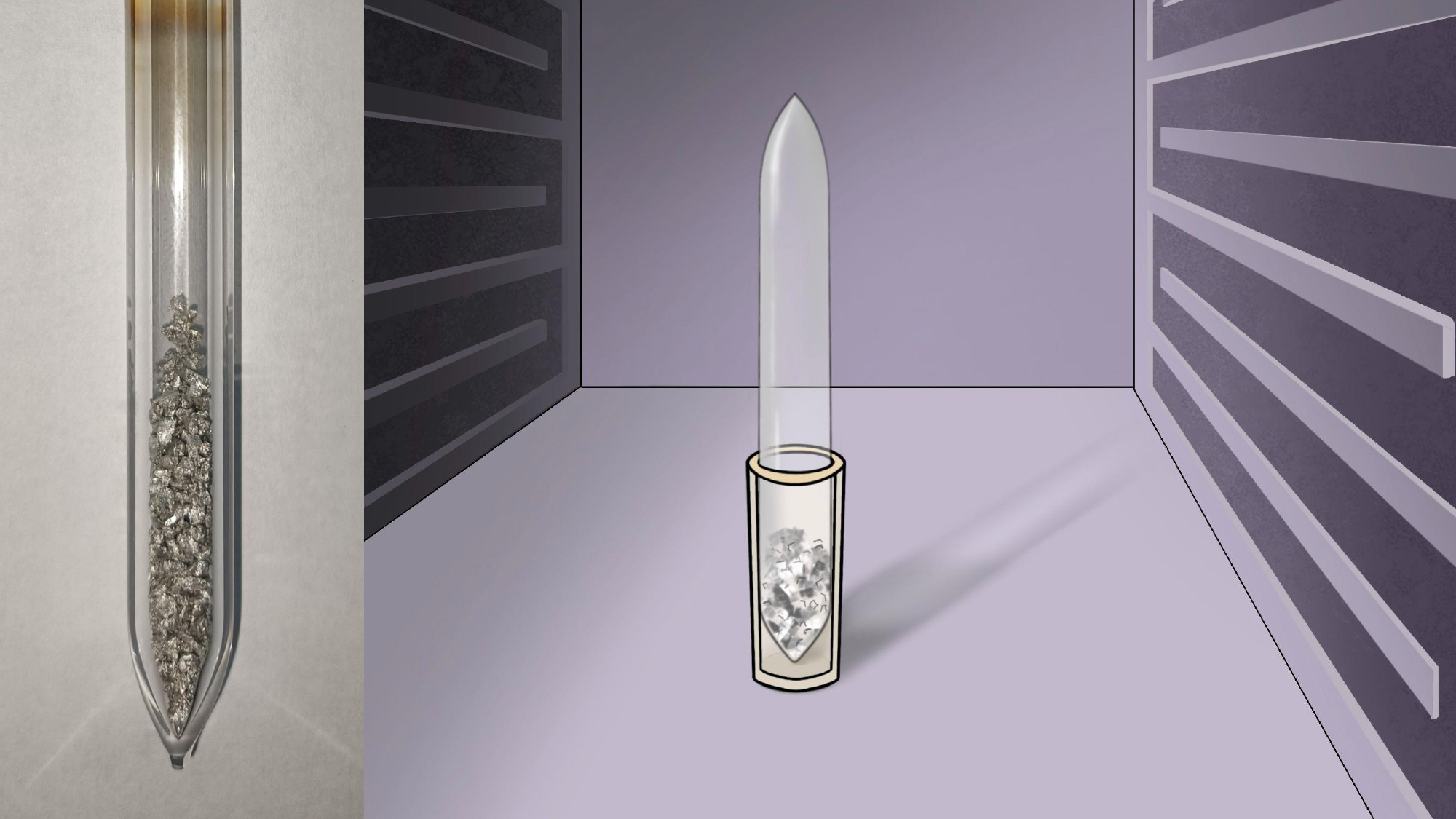}
\caption{\label{Fig:S1-SI} Photographic image and the schematic diagram of the vacuum sealed conical-end quartz ampule for the modified Bridgman growth. The schematic diagram depicts the quartz ampule kept vertically upright inside the box furnace.}
\end{figure}

\begin{figure}[htp]
\includegraphics[width=\columnwidth]{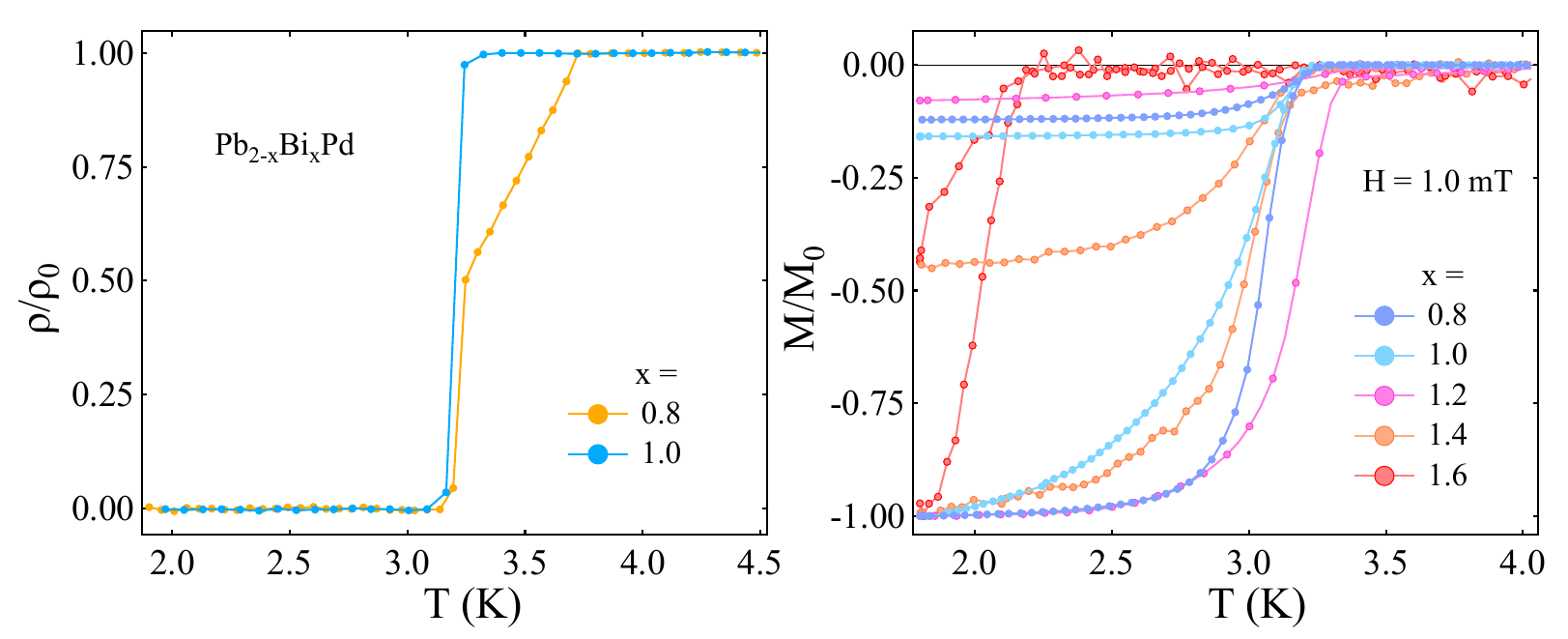}
\caption {\label{Fig:S2-SI} (a) Zero drop in resistivity for the mixed-phase bulk superconductors with $x=0.8$ and $1.0$. (b) Temperature dependence of the normalized magnetization in ZFCW and FCC modes under a 1.0 mT magnetic field for \ch{Pb_{2-x}Bi_{x}Pd}, $x=0.8,1.0,1.2,1.4,$ and $1.6$ single crystals.}
\end{figure}

\begin{table}[htp]
\begin{threeparttable}
\caption{The summary of variable growth conditions of \ch{Pb_{2-x}Bi_{x}Pd} ($0.2\le x\le1.8$) crystals using the modified Bridgman growth technique.}
\label{tbl: growth}
\setlength{\tabcolsep}{15pt}
\renewcommand{\arraystretch}{1.5} 
\begin{tabular}[b]{cccccc}\hline \hline
$x$ & \makecell{Ramping Rate \\ (K/hr)} & \makecell{Reaction Temp. \\ (K)} & \makecell{Dwelling Time \\ (hrs)} & \makecell{Cooling Rate \\ (K/hr)} & \makecell{Ice Quenching \\ Temp. (K)}\\
\hline
0.2 & 30 & 873 & 48 & 2.0 & 729\\
0.4 & 30 & 873 & 48 & 2.0 & 729\\
0.6 & 30 & 1073 & 48 & 2.0 & 729\\
0.8 & 30 & 1173 & 72 & 2.0 & 673\\
1.0 & 30 & 1273 & 72 & 2.0 & 753\\
1.2 & 30 & 1173 & 72 & 2.0 & 673\\
1.4 & 30 & 1173 & 48 & 2.0 & 673\\
1.6 & 30 & 1173 & 48 & 2.0 & 763$^{\dagger}$ \\
 &  &  & 12 & 0.5 & 693 \\
1.8 & 30 & 1173 & 48 & 2.0 & 873\\
\hline \hline
\end{tabular}
\begin{tablenotes}
    \small
    \item[$\dagger$]{The cycle is continued in the next line.}
\end{tablenotes}
\end{threeparttable}
\end{table}

\begin{figure}[ht]
\includegraphics[width=.75\columnwidth]{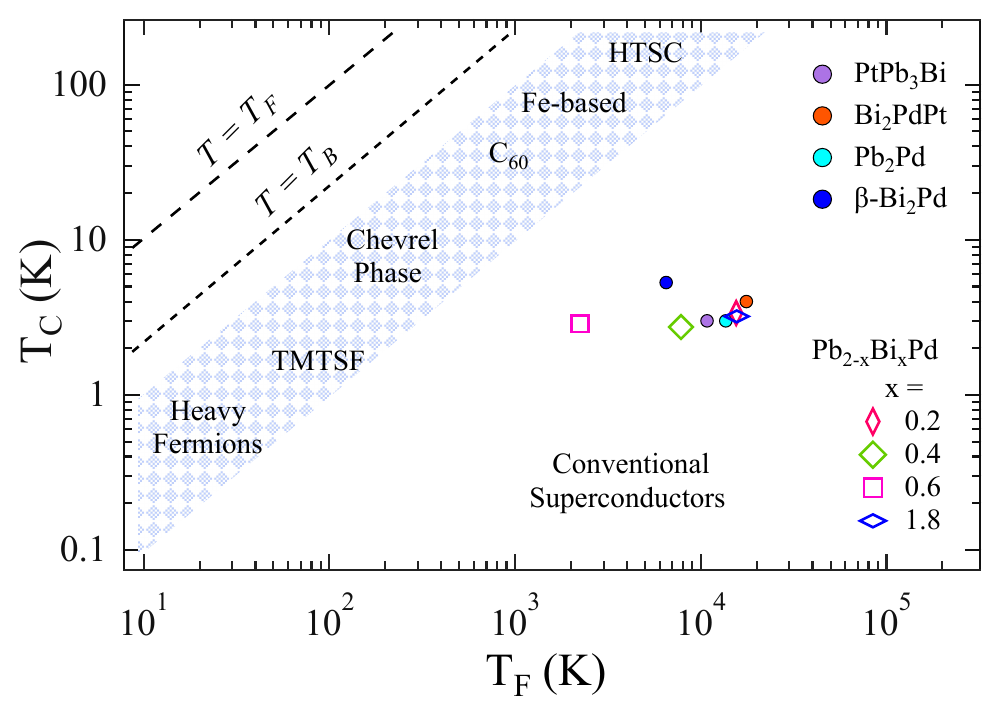}
\caption {\label{Fig:S3-SI} The Uemura plot, showing $T_C$ versus $T_F$ for \ch{Pb_{2-x}Bi_{x}Pd}, $x=0.2,0.4,0.6,$ and $1.8$ single crystals, compared with the parent compounds \ch{Pb2Pd} and $\beta$-\ch{Bi2Pd} other Pb, Bi, and Pd-based compounds.}
\end{figure}

\end{document}